\documentclass[a4paper]{report}
\usepackage[utf8]{inputenc}
\usepackage[T1]{fontenc}
\usepackage{RJournal}
\usepackage{amsmath,amssymb,array}
\usepackage{booktabs}

\begin{document}

\sectionhead{Contributed research article}
\volume{XX}
\volnumber{YY}
\year{20ZZ}
\month{AAAA}

\begin{article}
\title{Bayesian Inference for Multivariate Spatial Models with \pkg{R-INLA} }
\author{by Francisco Palm\'i-Perales, Virgilio G\'omez-Rubio, Roger S. Bivand, Michela Cameletti and H\aa{}vard Rue}

\maketitle

\abstract{
Bayesian methods and software for spatial data analysis are generally now well established in the scientific community. Despite the wide application of spatial models, the analysis of multivariate spatial data using \pkg{R-INLA} has not been widely described in the existing literature. Therefore, the main objective of this article is to demonstrate that \pkg{R-INLA} is a convenient toolbox to analyse different types of multivariate spatial datasets. Additionally, this will be illustrated by analysing three datasets which are publicly available. Furthermore, the details and the R code of these analyses are provided to exemplify how to adjust multivariate spatial datasets with \pkg{R-INLA}. 
}

\section{Introduction}
\label{Introduction}

\medskip
Multivariate spatial models have been studied by several authors. For example, \citet{VanLieshout1999} describe dependence between multivariate point patterns  by proposing novel summary statistics. Additionally, multivariate log-Gaussian Cox processes have been used to analyse multivariate point patterns \citep{DiggleMoraga, WaagerMultLogGau, gomez2015analysis}.
Furthermore, several studies have been published for analysing multivariate lattice data. For example, in \citet{macnab2018} an insight of the generalization of the univariate models to multivariate models is extensively discussed and in \citet{martinez2017} a framework to analyse multiple variables is proposed in the context of disease mapping. A review of the multivariate spatial models in disease mapping is performed in \citet{LibroMigue}. These studies use Markov Chain Monte Carlo \citep[MCMC, ][]{MCMC} methods to exemplify their proposals. In the context of spatial modeling, these methods can lead to a high computation cost. 

\medskip
In recent times, the integrated nested Laplace approximation \citep[INLA, ][]{INLA} has provided an alternative way of fitting Bayesian hierarchical models. Furthermore, when estimating continuous processes, the stochastic partial differential equation approach \citep[SPDE, ][]{SPDE} combined with INLA can be employed. The different spatial models which can be fit with \pkg{R-INLA} \citep{RINLA} have been compiled by several authors \citep{LindgrenRue2015, BlangiardoCameletti, Bakkaetal:2018}. For a recent review, the reader is referred to Chapter 7 of \citet{gomez2020bayesian}. \citet{Krainskietal:2019} provide an exhaustive tutorial about how to adjust advanced spatial models in \pkg{R-INLA}. Additionally, INLA can be combined with other algorithms such as MCMC techniques in order to adjust models which can not be adjusted solely with INLA \citep{GomezRubioPalmiPerales:2018}.

\medskip
The  aim of this paper is to provide a summary of how to analyze the different types of multivariate spatial data with \pkg{R-INLA}. The increasing interest (and application) of multivariate models have been the main motivation of this work. Therefore, the main objective of this article is to describe how to analyse any multivariate spatial dataset using \pkg{R-INLA}, which appears as an interesting toolbox to analyse this models' class in a Bayesian framework. 

\medskip
The remainder of the manuscript is structured as follows. First, a brief description of the different multivariate models is detailed in the case of areal, geostatistical and point pattern data. The next sections discuss the prior distribution choice and the structure the data. Then, three examples about how to analyse multivariate spatial models with \pkg{R-INLA} are detailed illustrating the main objectives. Finally, a brief summary of the conclusions of this work appear in the last part of this manuscript.

\section{Multivariate Spatial Models}
\label{MultModels}

\medskip
The R programming language offers a wide range of standalone packages for analysing spatial datasets. Several of them focus on a particular type of spatial data. For instance, point pattern analysis can be performed with \CRANpkg{spatstat} \citep{spatstat:2015} and \CRANpkg{spatialkernel} \citep{spatialkernel}. Geostatistical data can be modeled using \CRANpkg{gstat} \citep{gstat, gstatRpackage} or \CRANpkg{spBayes} \citep{spBayes, spBayes2}. Other R packages such as \CRANpkg{CARBayes} \citep{CARBayes} are designed to analyse lattice data.

\medskip
INLA is well suited to fit models with a multivariate response. In particular, multivariate models will consider a matrix response and each column is modeled using a different likelihood, but the linear predictors of these different likelihoods can share terms, so that dependence among the different response variables is introduced. In this section, a brief introduction of some multivariate spatial models is provided. These models will be fit with \pkg{R-INLA} in the examples.

\subsection{Areal data}
\label{sec2:lattice}

\medskip
When analysing lattice data, the domain is divided into non-overlapping areas in which the data are collected. It is usually considered that two areas are neighbours if they share a common boundary. This adjacency structure is often included to account for spatial autocorrelation \citep{banerjee2014hierarchical}. 

\medskip
When the value of several variables are recorded in each area the resulting data become a multivariate lattice dataset. The joint analysis of the spatial distribution of several variables allows to detect similar (spatial) patterns between some of these variables \citep[][ Chapter 10]{banerjee2014hierarchical} while estimating the spatial effects. We will illustrate the analysis of this type of data using a Poisson regression model commonly employed in spatial epidemiology to analyse count data but other similar models can be proposed for binary or continuous outcomes.

\medskip
Given the $d$-th variable  of interest (with $d=1,\ldots,D$) and area $i$ (with $i=1,\ldots,n$), the response of interest $Y_{d, i} $ can be modeled using a Poisson distribution with mean $\mu_{d,i}$:

$$
Y_{d, i} \sim \textrm{Po}(\mu_{d,i}) .
$$
\noindent
The mean is usually modeled as a sum of different terms through a link function. The selection of these terms depends on the available data and the proposed model structure. For instance, one option can be described as

$$
\psi(\mu_{d, i}) = \alpha_d + u_i + v_{d,i} 
$$ .

\noindent
Here, $\psi(\cdot)$ is a link function (which in this case is the logarithm function), $\alpha_d$ is a specific intercept, $u_i$ a shared (between all or a group of variables) spatial term and $v_{d,i}$ a variable-specific random effect. Note that restrictions may be required on $v_{d,i}$ to make all effects identifiable \citep{rueheld:2005}.


\medskip
In the context of disease mapping, the usual variables of interest are the counts of mortality or incidence of different diseases over the study region. Now $d$ represents the specific disease, therefore, following the above structure, the observed number of cases of the $d$th disease in the $i$ area, $Y_{d, i} $, can be modeled as

$$
Y_{d, i} \sim \textrm{Po}(\mu_{d,i} = E_{d,i}\cdot \theta_{d,i})
$$

$$
\log(\theta_{d,i}) = \alpha_d + u_i + v_{d,i}
$$

\noindent
where, $E_{d,i}$ and $\theta_{d,i}$ are the expected number of cases and the relative risk of disease $d$ in area $i$, respectively. As before, $\alpha_d$ is a disease specific intercept (to account for differences in the total number of observed cases), $u_i$ a shared spatial term (which does not depend on the disease) and $v_{d,i}$ a disease-specific spatial random effect. 

\medskip 
Several authors \citep[see, for example,][and the references therein]{MartinezBeneito:2013} have proposed different approaches to model multiple diseases in space and time. \citet{GomezRubioetal:2019} propose a separable spatio-temporal model with weighted shared components that can be used to detect diseases with similar patterns. In \citet{INLAMSM}, the authors have developed a R package (\CRANpkg{INLAMSM}) which builds on top on \pkg{R-INLA} to adjust some of the most common multivariate lattice approaches.

\subsection{Geostatistics}
\label{sec2:mgeo}

\medskip
Geostatistical datasets are built with values of variables which vary continuously over a spatial domain. These datasets contain observations which are geographically referenced, i. e. both the value and where it is collected (the coordinates) appear in the dataset. Then, the estimate of the variation surface of these variables is computed using geostatistical models.

\medskip
Similarly to lattice data, geostatistical multivariate models can be fit with INLA by sharing common terms. Here, the first variable can be modeled so that the mean includes a shared spatial term assumed to be a Gaussian process with a covariance defined using a Mat\'ern function, and all the other variables can depend on this shared spatial term plus specific spatial effects. Hence, for example, $K$ variables of interest ($Y_k$ for $k=0, \ldots, K-1$), with a general likelihood function $P(\cdot)$, measured at $n$ different locations can be written as: 

$$Y_{i,k} \sim P(\mu_{i,k})\\$$.

Then, the mean of the baseline variable ($\mu_{i,0}$) will be modelled as a sum of an intercept ($\alpha_0$) and a shared spatial effect ($u_{i,0}$). Furthermore, the mean of observation $i$ and variable $j$ ($\mu_{i,j}$) will be modeled through an intercept for each variable ($\alpha_j$), the shared spatial effect ($u_{i,0}$) and a specific spatial effect ($u_{i,j}$) as follows:

\begin{align}
\mu_{i,0} =& \alpha_0 + u_{i,0};  &i&=1,\ldots,n\nonumber\\
\mu_{i,j} =& \alpha_j + u_{i,0} + u_{i,j}  &i&=1,\ldots,n;\ j=1,\ldots,K-1\nonumber
\end{align}

\noindent
Here, $u_{i,0}$ represents the shared (between variables) spatial term, while  $u_{i,j}\, (j\geq1)$ represents specific terms that can be used to assess departures from the shared spatial term. These random effect terms are assumed to be Gaussian processes with covariance defined using a Matérn covariance function, which for a generic spatial random effect $u(s)$ is defined as:

$$Cov(u_p, u_l) = Cov(u(s_p), u(s_l))= \frac{\sigma^2}{\Gamma(\lambda)2^{\lambda-1}} (\kappa \|s_p - s_l\|)^{\lambda}K_{\lambda}(\kappa \|s_p - s_l\|)$$

\noindent
where $\|s_p - s_l\|$ represents the Euclidean distance between points $s_p$ and $s_l$ and $\sigma$ is the marginal variance of the latent Gaussian process. $\kappa$ is the scaling parameter, which is related to the range, and $K_{\lambda}$ represents the modified Bessel function of the second kind and order $\lambda$, which measure the smoothness of the process. Furthermore, this latent Gaussian field $u(s)$ can be approximated using the SPDE approach defined in \citet{SPDE} that relies on the finite element method through (an appropriate choose of) deterministic basis functions defined in a triangulation of the domain

$$u(s) =\sum_{k=1}^{m} \phi_k(s) w_k$$

\noindent
where $\phi_k(s)$ are the basis functions (pairwise linear functions), $m$ is the total number of nodes (triangle vertices) and $w_k$ are zero-mean Gaussian distributed weights. For more details, the reader is referred to \citet{SPDE}.

\medskip
In this example, it is considered that the different variables are measured at the same $n$ spatial points, but the measurements of each variable can be located in different locations of the study region leading in a misalignment framework. 

\medskip
A multivariate geostatistical analysis can be performed using the R package \pkg{gstat}. However, the models implemented in this R package are based on a classic and frequentist  statistical approach.
 
\subsection{Point patterns}
\label{sec2:mpp}

\medskip
A point pattern is defined as a group of points (geographically located) which are a single realization of a stochastic process called point process. A multivariate point pattern can be defined as a group of several point patterns where each point pattern has a different origin, i.e. each point pattern is caused by different processes. These are also referred to a specific case of \textit{marked} point pattern \citep{spatstat:2015} where each point pattern is labelled with a categorical mark.

For a completely random point process, points appear independently of each other and uniformly over the study region. This is also known as a homogeneous Poisson process with (constant) intensity $\lambda$, which measures the average number of points per unit area. It is also possible to consider a spatially varying intensity, $\lambda(x)$, so that the process becomes an \textit{inhomogeneous} Poisson process. Other types of point processes can be more complex \citep[sse, for example,][]{spatstat:2015}.

Several methods have been used to model the intensity function, $\lambda(s)$. Here, it will be considered as a continuous process over all the study region and it will be analysed using log-Gaussian Cox processes \citep{Moller,DiggleMoraga}. Log-Gaussian Cox processes can be fit by including spatial terms using the SPDE approach implemented in INLA \citep{Simpsonetal:2016}. The analysis of the intensity as a continuous function over the study region is similar to the case of multivariate geostatistics.

\medskip
Given $K$ point patterns in a region $D$, an example of how to structure a multivariate point patterns model is:

\begin{align*}
\log{\lambda_{0}(s)} =& \alpha_0 + u_{0}(s);  &s&\in D\\
\log{\lambda_{j}(s)} =& \alpha_j + u_{0}(s) + u_{j}(s)  &s&\in D;\ j=1,\ldots,K-1\nonumber
\end{align*}
\noindent
where $\lambda_{0}(s)$ is the intensity of the baseline point pattern and $\lambda_{j}(s)$ the intensity function of the $j$th point pattern. Moreover, $\alpha$ terms represent the intercepts and $u(s)$s are the spatial effects. Specifically, $u_0(s)$ is the shared (between the different point patterns) spatial term and $u_j(s)$ are the specific spatial terms which will catch the differences between each point pattern and the baseline one.

\medskip
In spatial epidemiology, researchers pursue to analyse whether a distribution of cases follows the spatial distribution of a set of controls, or it depends on exposure to pollution sources or other risk factors \citep{PalmiPerales2019Biometrical}. This is an application of the model described here, the log-intensity of the controls can be modeled using a shared spatial term and the log-intensity of the cases can include this shared spatial term plus a disease-specific spatial term. Furthermore, the linear predictor can include other terms to account for risk factors and exposure to pollution sources.

\section{Prior distribution selection}

\medskip
Prior choice is an essential step in a Bayesian inference process. In both, geostatistical and point patterns analysis, the SPDE approximation requires to set a prior distribution to the nominal range, $r$, and the nominal variance, $\sigma$. Penalized complexity prior distributions \citep[PC-priors, ][]{PCpriors} can be chosen for both parameters. In few words, PC priors are based on the idea of penalising the complexity from a baseline model, i.e., the prior density is related to the distance from a baseline model. A remarkable benefit of the PC-priors are their high intuitiveness in their definition. 

\medskip
In the case of point patters and geostatistical data, we establish a high probability of the range, $r$, been lower than the half of the maximum distance ($d_m$) of the study region, that is, $P(r < d_m/2 )$ is almost 1. In the case of the variance, following \citet{Simpson2019Careful}, an upper limit for the variation of the intensity ($U_{\alpha}$) is considered. Therefore, the probability of the standard deviation been greater than this upper limit ($ P(\sigma > U_{\alpha})$) is almost 0.  

\medskip
In the case of multivariate lattice data, a prior distribution should be assigned to the variance/standard deviation/precision of each spatial effect. Some authors have discussed the most appropriate vague prior distributions in these cases. For instance, \citet{GelmanHCpriors} proposes to avoid inverse Gamma distributions on the precision and propose some alternatives. In this article, flat uniform prior distributions are assigned to the standard deviation parameters.

\section{The structure of the multivariate spatial data in \pkg{R-INLA}}

\medskip
The spatial data sets must be formatted properly in order to be analysed with \pkg{R-INLA}. Their structure depends on the spatial data type. In this section, a brief explanation of how to tackle each dataset will be given. The following code will exemplify how to structure the spatial data in practice. 

\subsection{Areal data}

\medskip
Areal datasets usually are structured as a matrix where each row corresponds to a single area and each column corresponds to a single variable such as the number of cases of a specific disease. However, this is not the appropriate structure to analyse lattice data in \pkg{R-INLA}. 

Consider a dataset with $D$ variables measured in a study region divided in $n$ areas. In order to analyse this dataset with \pkg{R-INLA}, a matrix with $D$ columns and $D x n$ rows has to be built. Specifically, this matrix will store the $n$ values of the first variable ($D=1$) in the first column between the first and the $n$th row, then the following values of this column will be $NA$. In R, $NA$ represent an empty space. 

Then, the data of the second variable ($D=2$) would be placed in the second column and between the $n+1$ and the $2n$ rows. The rest of the values of the second column will be filled with $NA$s. Therefore, the filling of the rest of the matrix will be done following this procedure. 

As a toy example, let's consider $n=2$ and $D=3$, the original dataset is structured as a $2x3$ matrix: 

$$
\begin{bmatrix}
1 & 4 & 3 \\
2 & 6 & 5 
\end{bmatrix}
$$.

Following, the above procedure, the matrix that should be passed to \pkg{R-INLA} would be the following: 

$$
\begin{bmatrix}
1  & NA & NA \\
2  & NA & NA \\
NA & 4  & NA \\
NA & 6  & NA \\
NA & NA & 3 \\
NA & NA & 5 
\end{bmatrix}
$$

\subsection{Geostatistics}

\medskip
Geostatistical datasets contain the values of the different variables and the locations where they are measured. In this case, the main objective is to estimate the value of the variables as a continuous surface over the study region. The SPDE \citep{SPDE} will be used in order to analyse the multivariate geostatistical data using a discretization (a mesh) of the surface. 

\medskip
Following the SPDE approach, data must be \textit{stacked} using the appropriate format. The helper function \texttt{inla.stack()} can be used in order to built a stack object which contains the data, the effects considered in the model and the A matrix which define the location of the data on the mesh, that is, A shows which triangle of the mesh contains each data point. Specifically, a stack object will be created for each variable, then, the stack function will be used once more to combine all the stacks in a definitive stack object. 

\medskip
The measured values of each variable will be included in its stack using a vector which will follow the same procedure of the above subsection. As a general example, let's consider $n$ measures and $K$ variables. In this case, all the vectors will have $nxK$ length and, for instance, the vector of the stack of the first variable will store the $n$ values of the first component and the rest will be filled with $NA$s.  

\medskip
As a toy example, let's consider $n=2$ and $K=3$, and the values of the three variables which are stored in this matrix:

$$
\begin{bmatrix}
1.2 & 4.8 & 3.7 \\
2.1 & 6.5 & 5.4 
\end{bmatrix}
$$

The three vectors (each will be placed in the corresponding stack) are: $$[1.2, 2.1, NA, NA, NA, NA]^{\top}$$ $$[NA, NA, 4.8, 6.5, NA, NA]^{\top}$$ $$[NA, NA, NA, NA, 3.7, 5.4]^{\top}$$

\subsection{Point patterns}

\medskip
Multivariate point patterns datasets contain the locations of different point patterns. Here, the objective is to estimate the intensity surface of each point pattern. Hence, the SPDE is once more required. Therefore, the same procedure using the \texttt{inla.stack()} R function will be followed in this case. However, there is a special detail that has to be carefully tackled. 

\medskip
In the case of point patterns, the data included in the stack function have a specific structure. In this case, a list with two elements will be included for storing the data of each point pattern. 

\medskip
The first element of the list will be a matrix with $N_v + N_i$ rows and $K$ columns where $N_i$ are the total number of points of the $i$th point pattern, $K$ is the number of different point patterns and $N_v$ is the number of points (vertices) of the mesh (triangulation) needed in the SPDE approach. Then, the matrix of the stack of the $i$th point pattern will be filled with $NA$s except for the $i$th column. This $i$th column will contain firstly $N_v$ zeros corresponding with the point of the mesh. After these zeros, $N_i$ ones related with the $N_i$ points of the $i$th point pattern will be placed. 

\medskip
The second element of the list will be a vector containing an offset. Specifically, the length of this vector is also $N_v + N_i$ where the first $N_v$ elements will contain the "weights" of the mesh points. The reader is referred to \citet{Simpsonetal:2016} for more details about this approximation. Then, the rest of point will be zeros. 

\medskip
This list has to been built for each stack of the each point pattern. Then, as before, all the single stacks are combined in a single stack object. As a toy example, consider a mesh with $N_v =3$, two point patterns $K=2$ with three and four elements, respectively ($N_1=3$ and $N_2=4$). Then the data passed to the second stack is a list with these two elements:

$$ 
\begin{bmatrix}
NA & 0  \\
NA & 0  \\
NA & 0  \\
NA & 1  \\
NA & 1  \\
NA & 1  \\
NA & 1  
\end{bmatrix}
$$
$$ [2.3, 4.3, 6.2, 0, 0, 0, 0]^{\top}$$

\section{Examples}
\label{sec:Examp}

\subsection{Multivariate lattice data}

\cite{GomezRubioPalmiPerales:2018} study the spatial risk variation of three types of cancer in peninsular Spain. In particular, they consider oral cavity, esophagus and stomach cancer at the province level. In order to assess similar spatial variation, that may lead to the identification of shared risk factors, models with shared and disease specific spatial patterns can be proposed.

Data are available in a \code{RData} file available from GitHub \cite[see][for details]{GomezRubioPalmiPerales:2018}. The following code creates the response variable (by staking the three vectors of observed cases) as well as the expected counts (other variables are needed but not shown here). Note that due to the different structure of the variables involved, data are stored together in a \code{list} object instead of a \code{data.frame}. 

\begin{example*}
# Load data
load("dismap_sim_data.RData")

# Set shorter names
names(OyE.sim)[1:6] <- c("Obs.Cb", "Esp.Cb", "Obs.Eso",
  "Esp.Eso", "Obs.Est", "Esp.Est")

# Create a dataset for INLA (n x 3)
n <- nrow(OyE.sim)
d <- list(OBS = matrix(NA, nrow = n*3, ncol = 3))

# Add observed
d$OBS[1:n, 1] <- OyE.sim$Obs.Cb #Bucal cancer
d$OBS[n + 1:n, 2] <- OyE.sim$Obs.Eso #Esophagous cancer
d$OBS[2*n + 1:n, 3] <- OyE.sim$Obs.Est #Stomach cancer

# Expected cases
d$EXP <- c(OyE.sim$Esp.Cb, OyE.sim$Esp.Eso, OyE.sim$Esp.Est)
\end{example*}

As an example, we will consider a model in which the log-relative risk of oral cavity cancer ($\theta_{o,i}$) is modeled using an intercept ($\alpha_{o}$) plus a shared spatial term ($u_i$) using an ICAR specification following the model of the areal or lattice data analysis section. Furthermore, log-relative risks of esophagus ($\theta_{e,i}$) and stomach cancer ($\theta_{s,i}$) are modeled using disease-specific intercepts ($\alpha_e$ and $\alpha_s$) plus the shared spatial term and cancer-specific ICAR spatial terms ($v_{e,i}$ and $v_{s,i}$). Specifically, the relative risks are measured as follows: 

\begin{align*}
\log(\theta_{o,i}) &= \alpha_{o} + u_i  \\
\log(\theta_{e,i})  &= \alpha_e + u_i + v_{e,i}\\
\log(\theta_{s,i})  &= \alpha_s + u_i + v_{s,i}  \, \,\, i=1,\ldots, n\nonumber
\end{align*}

These spatial specific terms can be used to assess departures from the shared spatial term. The chosen prior distributions for the standard deviation of all the effects are flat uniform distributions.

The model formula is defined in the model below. The \code{rf} object captures the value of the intercepts (the value of the above $\alpha$s ). Latent effects of type \code{"copy"} are used to define shared terms  in the model. Furthermore, some latent effects have the uniform prior for the standard deviation defined in object \code{prior.prec}.

\begin{example*}
# Formulas for the model
form <- OBS ~ -1 + rf +
  f(copy1, model = "besag", graph = W, hyper = list(prec = prior.prec)) +
  f(copy2, copy = "copy1", fixed = TRUE) +
  f(copy3, copy = "copy1", fixed = TRUE) +
  f(spatial2, model = "besag", graph = W, hyper = list(prec = prior.prec)) +
  f(spatial3, model = "besag", graph = W, hyper = list(prec = prior.prec))
\end{example*}

Finally the model is fit with \pkg{R-INLA} using the code below. Note how the \code{family} argument takes a vector of three elements as this likelihood has three components (one for each disease).

\begin{example*}
res <- inla(formula =  form, data = d, family = rep("poisson", 3), E = d$EXP)
\end{example*}

\begin{figure}[h!]
\begin{center}
\includegraphics[scale=0.48]{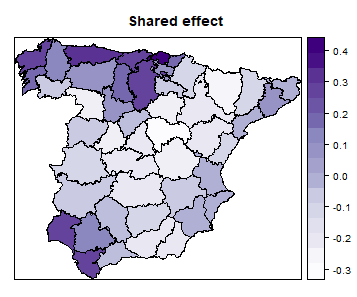}
\includegraphics[scale=0.48]{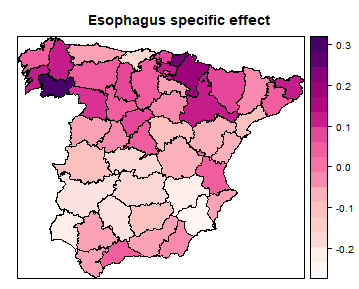}
\includegraphics[scale=0.48]{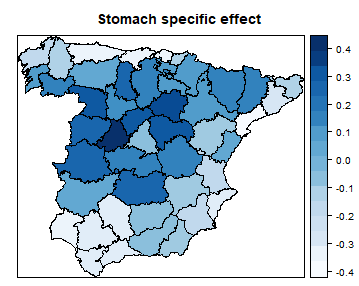}
\end{center}
\caption{Posterior mean of the shared spatial pattern (left), the esophagus-specific (middle) and stomach-specific (right) spatial patterns.}
\label{fig:spain}
\end{figure}

Figure~\ref{fig:spain} shows the different spatial terms in the model. The shared specific term represents the spatial variation of the risk of oral cavity cancer and also serves as a baseline for the other types of cancer. The esophagus specific spatial term is quite mild, which indicates that these two types of cancer have a very similar spatial pattern. The specific spatial pattern for stomach cancer shows some provinces in the center of the country with a higher mortality than oral cavity (and esophagus).

\subsection{Multivariate geostatistics}

The \code{meuse} dataset in the \pkg{gstat} package gives the locations and measurements of topsoil heavy metals collected in a flood plain by the Meuse river, close to the village of Stein (The Netherlands). After loading the \pkg{gstat} package, the data can be loaded using the following code:

\begin{example*}
# Load the data
data(meuse)

# Create the spatial object
coordinates(meuse) = ~x+y
proj4string(meuse) <- CRS("+init=epsg:28992")
\end{example*}

These measurements are highly correlated and we will explore in this example how to fit geostatistical models with the SPDE approach. Note that now observations do not need to be in a regular grid. Instead of a grid, a mesh is defined to apply the SPDE approach. The boundary of the study region is stored in object \code{meuse.bdy} (see accompanying code). The definition of the mesh is done using the coordinates of the boundary of the area following the code below:

\begin{example*}
# Create the mesh
mesh <- inla.mesh.2d(boundary = meuse.bdy, loc = coordinates(meuse),
  max.edge = c(250, 500), offset = c(250, 500), n=c(32, 32))
\end{example*}

\begin{figure}[h!]
\begin{center}
\includegraphics[scale=0.5]{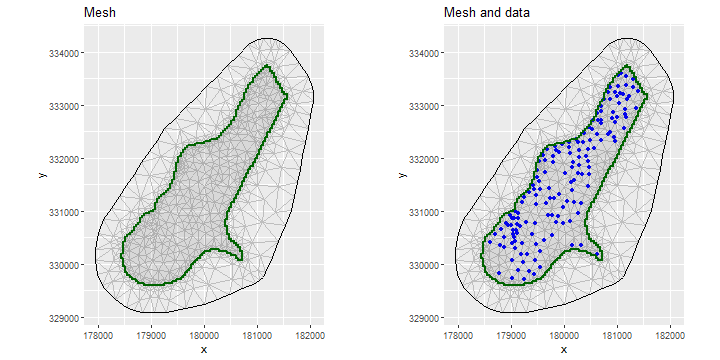}
\end{center}
\caption{Mesh used in the estimation of the concentration of heavy metals around river Meuse, only with the boundary of the study region (left) and jointly with the survey locations (right).}
\label{fig:meuse:mesh}
\end{figure}

The left plot of Figure~\ref{fig:meuse:mesh} shows the mesh built with the above code for this example.

In particular, the model will consider concentrations of lead and zinc. This concentrations have been measured in diferent locations which are displayed in the right part of Figure \ref{fig:meuse:mesh}. However, more heavy metals can be analysed following the same structure.  Values are considered in the log-scale. These log-transformed concentrations ($y_{l}$ and
$y_z$) are assumed to be normally distributed. The mean of the concentration of
lead is modeled using an intercept $\alpha_l$ plus a Gaussian process with a
Matérn correlation, $u_{i,s}$, while the mean of the concentration of zinc is
modeled using an intercept, $\alpha_z$ (different from the previous one), the
shared spatial effect, $u_{i,s}$, plus another spatial Gaussian process with
Matérn covariance, $u_{i,z}$. This will allow us to assess differences in the
spatial distribution of the concentration of both heavy metals. 

Following the geostatistics subsection of the multivariate spatial models section, the model can be written as: 

$$\log{y_{l}} \sim N(\mu_{i,l}, \sigma_l)\\$$
$$\log{y_{z}} \sim N(\mu_{i,z}, \sigma_z)\\$$

\noindent
where $\mu_{i,l}$ and $\mu_{i,z}$ represent the mean of the concentration of lead and zinc, respectively and which are modelled as follows:

\begin{align}
\mu_{i,l} =& \alpha_l + u_{i,s};  &i&=1,\ldots,n\nonumber\\
\mu_{i,z} =& \alpha_z + u_{i,s} + u_{i,z}  &i&=1,\ldots,n\nonumber
\end{align}

In this example, the prior choices are: 

$$ P(r < 2394.16) = 0.95  $$
$$ P(\sigma > 1000) = 0.05 $$

\noindent
where a nominal range higher than the half of the maximum distance (i.e., 2394.16 meters) of the domain is unlikely. Similarly for the nominal standard deviation, which is really unlikely to be higher than 1000 mg/kg of soil (ppm). These prior distributions are specified using the \code{inla.spde2.pcmatern()} function:

\begin{example*}
spde <- inla.spde2.pcmatern(mesh = mesh,
  prior.range = c(2394.16, 0.95), prior.sigma = c(1000, 0.05)) 
\end{example*}

The SPDE approximation estimates the spatial random effects at the vertices
of the mesh, so that estimates at any other point are based on the estimates at the vertices of the triangle that contains the point. The position of this point inside the triangle is identified
using barycentric coordinates \citep[see][for details]{Krainskietal:2019}.
The projector matrix contains all these coordinates for all the points in the
dataset and it is obtained with function \code{inla.spde.make.A()}:

\begin{example*}
A.m <- inla.spde.make.A(mesh = mesh, loc = coordinates(meuse))
\end{example*}

Then, the data are structured using the \code{inla.stack()} function specifying three elements: the data, the projector matrix and the different effects of the linear predictor. Note that the data are prepared in the first two lines, then the \code{stack} objects are built. 

\begin{example*}
# Prepare the stack data
y1 <- log(meuse@data["lead"]); names(y1) <- "y.1"
y2 <- log(meuse@data["zinc"]); names(y2) <- "y.2"

# Create the stack object for cadmium
stk.lead <- inla.stack(
  data = list(log.y = cbind(y1[,1], NA)),
  A = list(A.m, 1),
  effects = list(spatial.field.lead = 1:spde$n.spde,
    data.frame(Intercept.lead = 1, dist.lead = meuse$dist)),
  tag = "Lead")

# Create the stack object for zinc
stk.zinc <- inla.stack(
  data = list(log.y = cbind(NA, as.vector(y2[, 1]))),
  A = list(A.m, A.m, 1),
  effects = list(
    spatial.field.zinc = 1:spde$n.spde, base.copy.zinc = 1:nv,
    data.frame(Intercept.zinc = 1, dist.zinc = meuse$dist)),
  tag = "Zinc")
\end{example*}

A projector matrix and a stack for the prediction grid have been also created following the below lines: 

\begin{example*}
# Create the projector matrix for the prediction
A.pr <- inla.spde.make.A(mesh = mesh, loc = coordinates(meuse.grid))

# Prepare the data for the prediction
y5 <- matrix(NA, ncol = 1, nrow = nrow(meuse.grid) ) 
y6 <- matrix(NA, ncol = 1, nrow = nrow(meuse.grid) )

# Build predicting stack for cadmium
stk.lead.pr <- inla.stack(
  data = list(log.y = cbind(as.vector(y5[, 1]), NA)),
  A = list(A.pr, 1),
  effects = list(spatial.field.lead = 1:spde$n.spde,
    data.frame(Intercept.lead = 1, dist.lead = meuse.grid$dist)),
  tag = "Lead.pred")

# Build predicting stack for zinc
stk.zinc.pr <- inla.stack(
  data = list(log.y = cbind(NA, as.vector(y6[, 1]))),
  A = list(A.pr, A.pr, 1),
  effects = list(
    spatial.field.zinc = 1:spde$n.spde, base.copy.zinc = 1:nv,
    data.frame(Intercept.zinc = 1, dist.zinc = meuse.grid$dist)),
  tag = "Zinc.pred")

\end{example*}

In this example, a \code{stack} object is built for the specific and the shared effect in order to study the spatial trend of this effects: 

\begin{example*}

# Stack for the shared effect
stk.shared <- inla.stack(
  data = list(log.y = cbind(as.vector(y5[, 1]), NA)),
  A = list(A.pr),
  effects = list(spatial.field.lead = 1:spde$n.spde),
  tag = "Shared")

# Stack for the specific sp effect zinc
stk.zinc.spec <- inla.stack(
  data = list(log.y = cbind(NA, as.vector(y6[, 1]))),
  A = list(A.pr),
  effects = list(spatial.field.zinc = 1:spde$n.spde),
  tag = "Zinc.spec")
\end{example*}

All the \code{stack} objects are pulled together in a single joint \code{stack} object using the \code{inla.stack()} function: 

\begin{example*}
# Put all the stacks together
join.stack <- inla.stack(
  stk.lead, stk.zinc, 
  stk.zinc.pr, stk.lead.pr,
  stk.shared, stk.zinc.spec)
\end{example*}

The model formula is defined below. The latent effect of type \code{"copy"} is used to define the shared term of the model. 

\begin{example*}
# Formulas for the model
form <- log.y ~ -1 + Intercept.lead + Intercept.zinc + dist.lead + dist.zinc + 
  f(spatial.field.lead, model = spde) +
  f(spatial.field.zinc, model = spde) + 
  f(base.copy.zinc, copy = "spatial.field.lead", fixed = TRUE)  

\end{example*}

Finally the model is fit with \pkg{R-INLA}. Note how the \code{family} argument takes a vector of two \code{"gaussian"} elements (one for each heavy metal concentration). Furthermore, note that the data are obtained from the joint stack with the \code{inla.stack.data()} function. 

\begin{example*}
meuse.res <- inla(formula = form, verbose = FALSE, 
  data = inla.stack.data(join.stack, spde = spde),
  family = rep("gaussian", 2), 
  control.family = list(zero.prec, zero.prec),
  control.predictor = list(A = inla.stack.A(join.stack), compute = TRUE),
  control.compute = list(dic = TRUE, waic = TRUE, cpo = TRUE, mlik = TRUE, po = TRUE))
\end{example*}

\begin{figure}[h!]
\begin{center}
\includegraphics[scale=0.35]{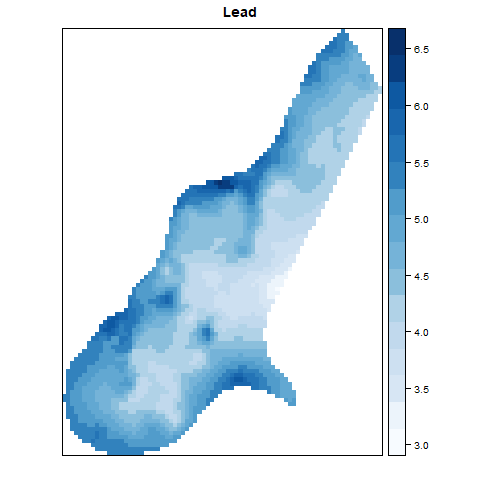}
\includegraphics[scale=0.35]{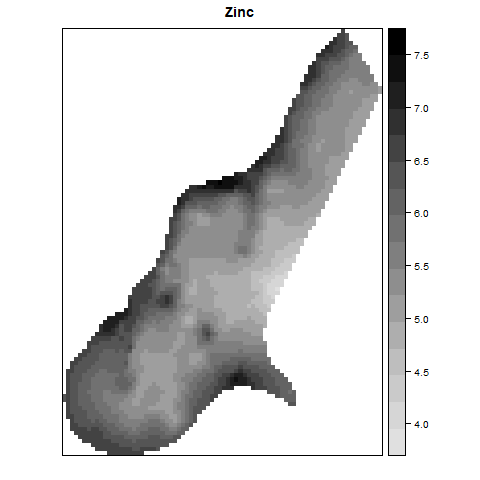}
\includegraphics[scale=0.35]{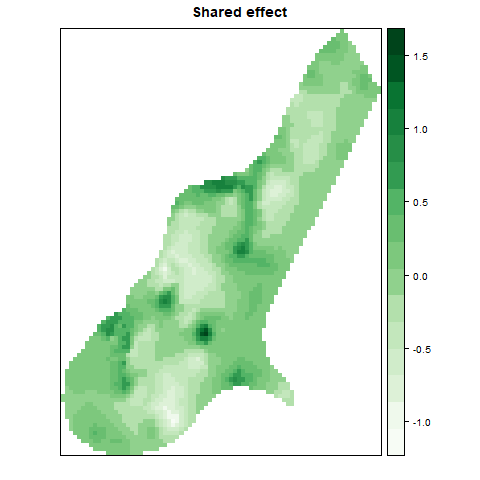}
\includegraphics[scale=0.35]{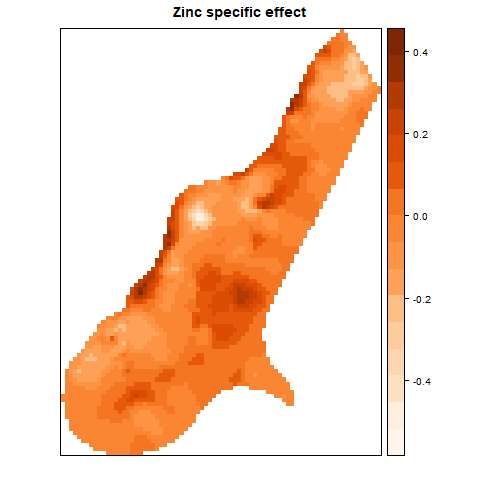}
\end{center}
\caption{Estimate of the posterior mean of the log-concentration of lead (top-left) and zinc (top-right). Estimates of the posterior mean of the shared spatial effect (bottom-left) and the zinc-specific spatial effect (bottom-right).}
\label{fig:meuse:est}
\end{figure}

Figure~\ref{fig:meuse:est} shows the estimates (posterior means) of the log-concentration of lead and zinc; the posterior mean of the shared and the zinc-specific effect are also shown. Note the strongly similar spatial pattern between both estimates.

\subsection{Multivariate point patterns}
\label{exam:mpp}

\begin{figure}[h!]
\begin{center}
\includegraphics[scale=0.40]{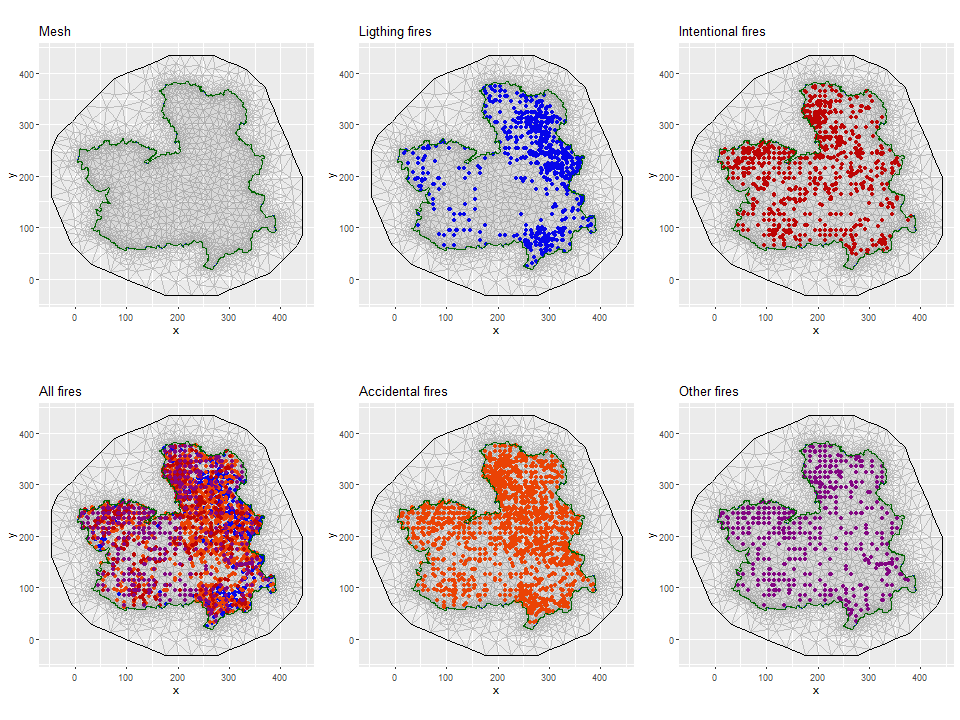}
\end{center}
\caption{The mesh used is shown alone (top-left) and with all the fire types (bottom-left). Furthermore, the mesh is also displayed for each fire type separately: lightning fires (top-middle), accidental fires (bottom-middle), intentional (top-right) and other fires (bottom-right).}
\label{fig:clmfires:data}
\end{figure}

\medskip
The \pkg{spatstat} package contains the \code{clmfires} dataset. This dataset records the occurrence of forest fires in the region of Castilla-La Mancha (Spain) from 1998 to 2007. These forest fires are classified by four different causes: lightning, accidental, intentional and other fires (Figure \ref{fig:clmfires:data}). After loading the packages, the data are obtained using the \code{data()} function:

\begin{example*}
#Load and display the data
data("clmfires")
\end{example*}

Specifically, in this example, the intensity of all of the types of forest fires is estimated by considering lightning fires as the baseline pattern. Furthermore, INLA is used to assess the similarities or differences between their spatial patterns. First of all, the mesh is built using the coordinates of the boundary \code{bdy.SP} of the dataset by following the below code. 

\begin{example*}
mesh <- inla.mesh.2d(
  boundary = list(bdy.SP, NULL), cutoff = 2, max.edge = c(20, 50),
  min.angle = 27, offset = c(1, 50), n=c(16,16))
\end{example*}

Following the model structure detailed in the point patterns subsection of the multivariate models section, the log-intensity of the lightning fires $\lambda_l(s)$ will be modeled using an intercept, $\alpha_l$, and a spatial Gaussian effect with Mat\'ern covariance, $u_0(s)$ as follows 

$$\log{\lambda_l(s)}= \alpha + u_0(s); \,\,\, s\in D$$  

Spatial effect $u_0(s)$ will also be shared in the linear predictor of the other types of fire. Similarly, the log-intensity of the accidental, intentional and the other fires (with unknown cause) will be modeled using specific intercepts, plus the shared spatial effect plus a specific spatial effect as follows: 

$$\log{\lambda_a(s)}= \alpha_a + u_0(s) + u_a(s); \,\,\, s\in D $$ 
$$\log{\lambda_i(s)}= \alpha_i + u_0(s) + u_i(s); \,\,\, s\in D $$  
$$\log{\lambda_o(s)}= \alpha_o + u_0(s) + u_o(s) \,\,\, s\in D $$  
 
\noindent
where these disease specific spatial effects can be used to assess any deviation from the shared pattern. Following the prior specification section, the chosen PC-priors in this example are:

$$ P(r < 200) = 0.95$$
$$ P(\sigma > 100) = 0.05 $$

\noindent
where a nominal range higher than half the maximum distance (e.g., 200 kilometres) of the domain is unlikely. Similarly for the nominal standard deviation, which is really unlikely to be higher than 100 fires per unit area in this context. These prior distributions have been specified using the function \code{inla.spde2.pcmatern()}:

\begin{example*}
spde <- inla.spde2.pcmatern(mesh = mesh,
  prior.range = c(200, 0.95),  prior.sigma = c(10, 0.05)) 
\end{example*}

 where the argument \code{prior.range} sets the prior for the range and the argument \code{prior.sigma} sets the prior for the standard deviation of the spatial effect. Then, the \code{stack} objects are built.

As stated above, when analysing point patterns it is necessary to assign some weights to the observed points. This is done by creating a Voronoi tesselation using these points, so that the area of the associated polygon becomes the associated weight. The following code illustrates how to obtain the Voronoi tesselation
and associated weights:
 
\begin{example*}
require(deldir)
dd <- deldir(mesh$loc[, 1],mesh$loc[, 2])
# Create a list of tiles in a tessellation
mytiles <- tile.list(dd)

if (!require(gpclib)) install.packages("gpclib", type = "source")
pl.study <- as(bdy, "gpc.poly") # Class for polygons
area.poly(pl.study) # Computing the area of the whole polygon

# Compute weight as area of the polygon given as an
# interaction between Voronoi tiles and domain polygon
w <- unlist(lapply(mytiles,
  function(p) area.poly(
    intersect(as(cbind(p$x,p$y), "gpc.poly"), pl.study)
    )
)
)
\end{example*}

These computed weights are introduced as the expected weights on the mesh points. Furthermore, the data should be structured as follows

\begin{example*}
# Data for the stack function: lighting fires
e.lig <- c(w, rep(0, n.lig))
y.lig <- matrix(NA, nrow = nv + n.lig, ncol = n.pp)
y.lig[, 1] <- rep(0:1, c(nv, n.lig))

# Data for the stack function: accidental fires
e.acc <- c(w, rep(0, n.acc))
y.acc <- matrix(NA, nrow = nv + n.acc, ncol = n.pp)
y.acc[, 2] <- rep(0:1, c(nv, n.acc))

# Data for the stack function: intentional fires
e.int <- c(w, rep(0, n.int))
y.int <- matrix(NA, nrow = nv + n.int, ncol = n.pp)
y.int[, 3] <- rep(0:1, c(nv, n.int))

# Data for the stack function: other fires
e.oth <- c(w, rep(0, n.oth))
y.oth <- matrix(NA, nrow = nv + n.oth, ncol = n.pp)
y.oth[, 4] <- rep(0:1, c(nv, n.oth))
\end{example*}

Another element of the SPDE approach is the projector matrix. When analysing multivariate point patterns this matrix has two parts: one for the mesh points (named \code{imat}) and the other for the points of the point pattern (named \code{lmat}). Then, the projector matrix is the combination of these two matrices:

\begin{example*}
#imat: define imat
imat <- Diagonal(nv, rep(1, nv))

#lmat: define lmat
lmat.lig <- inla.spde.make.A(mesh, pts.lig)
lmat.acc <- inla.spde.make.A(mesh, pts.acc)
lmat.int <- inla.spde.make.A(mesh, pts.int)
lmat.oth <- inla.spde.make.A(mesh, pts.oth)

#Projector matrix: Put together imat and lmat
A.lig <- rbind(imat, lmat.lig)
A.acc <- rbind(imat, lmat.acc)
A.int <- rbind(imat, lmat.int)
A.oth <- rbind(imat, lmat.oth)
\end{example*}

Once all the elements of the \code{stack} object are set, the \code{inla.stack()} function is used to built  the different \code{stack} objects. For instance, the \code{stack} objects of two (out of four) types of forest fires are shown

\begin{example*} 
# Create the stack for the lighting fires
stk.lig <- inla.stack(
  data = list(y = y.lig, e = e.lig),
  A = list(A.lig, 1),
  effects = list(spatial.field.lig = s.index.lig, 
    data.frame(Intercept.lig = rep(1, dim(A.lig)[1])) 
    ),
  tag = "Lighting")

# Create the stack for the accidental fires
stk.acc <- inla.stack(
  data = list(y = y.acc, e = e.acc), 
  A = list(A.acc, A.acc, 1), 
  effects = list(
    base.copy.acc = 1:nv, 
    spatial.field.acc = s.index.acc, 
    data.frame(Intercept.acc = rep(1, dim(A.acc)[1]))
    ),
  tag = "Accidental")
\end{example*}

As in the geostatistical example, the \code{stack} objects for the predictions have to be built following the same structure. Also, all the \code{stack} objects are joined in a single joint \code{stack} object as follows: 

\begin{example*}

# All stacks together
join.stack <- inla.stack(
  stk.lig, stk.acc, stk.int, stk.oth, 
  stk.lig.pr, stk.acc.pr, stk.int.pr, stk.oth.pr, 
  stk.shared, stk.acc.spec, stk.int.spec, stk.oth.spec)

\end{example*}

Note that in this example a \code{stack} object has been built not only for each forest fire types' linear predictor but also for the shared and the three different specific spatial effects. As in the other two examples, the model formula is defined in the model below. Latent effects of type \code{"copy"} are used to define the shared terms of the model. 

\begin{example*}
form <- y ~ -1 + Intercept.lig + Intercept.acc + Intercept.int + Intercept.oth +
  f(spatial.field.lig, model = spde) +
  f(spatial.field.acc, model = spde) + 
  f(base.copy.acc, copy = "spatial.field.lig", fixed = TRUE) +
  f(spatial.field.int, model = spde) + 
  f(base.copy.int, copy = "spatial.field.lig", fixed = TRUE) +
  f(spatial.field.oth, model = spde) + 
  f(base.copy.oth, copy = "spatial.field.lig", fixed = TRUE)    

\end{example*}

Finally the model is fit with \pkg{R-INLA} using the code below. Note how the \code{family} argument takes a vector of four Poisson elements (one for each fire type). Furthermore, note that the data are obtained from the joint stack with the \code{inla.stack.data()} function. 

\begin{example*}
pp.res <- inla(formula=form, verbose = FALSE, 
  data = inla.stack.data(join.stack, spde = spde), 
  family = rep("poisson", 4), 
  control.predictor = list(A = inla.stack.A(join.stack), compute = TRUE, link = 1),
  control.compute = list(dic = TRUE, waic = TRUE, cpo = TRUE, mlik = TRUE, po = TRUE)
)
\end{example*}

\begin{figure}[h!]
\begin{center}
\includegraphics[scale=0.40]{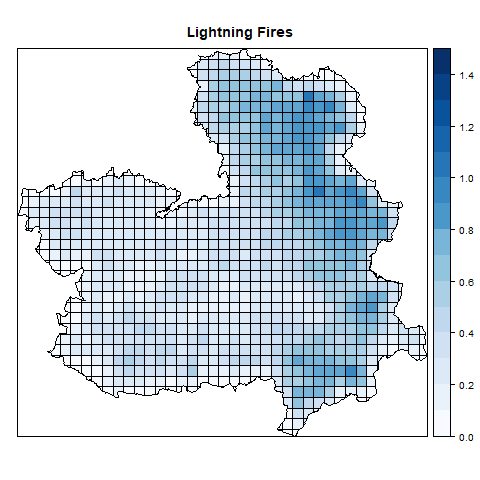}
\includegraphics[scale=0.40]{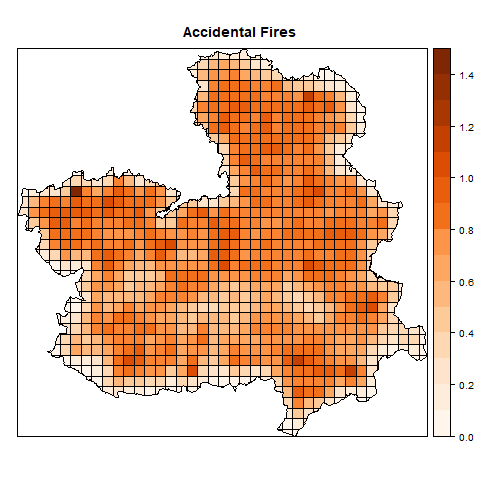}
\includegraphics[scale=0.40]{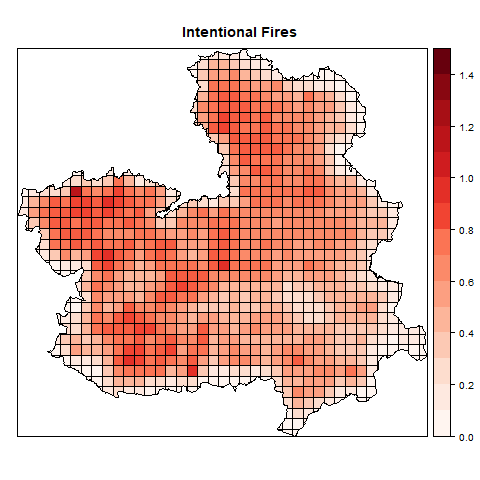}
\includegraphics[scale=0.40]{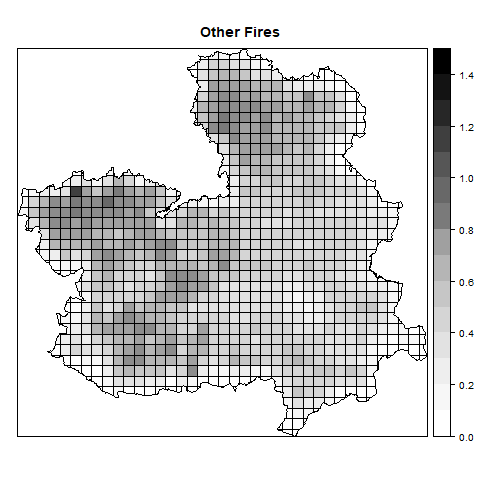}
\end{center}
\caption{Posterior mean of the log-intensity of the lightning (top-left), accidental (top-right), intentional (bottom-left) and other (bottom-right) fires.}
\label{fig:clmfires:est1}
\end{figure}

\begin{figure}[h!]
\begin{center}
\includegraphics[scale=0.40]{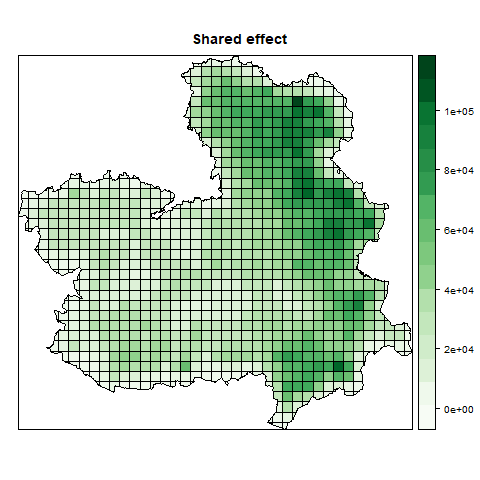}
\includegraphics[scale=0.40]{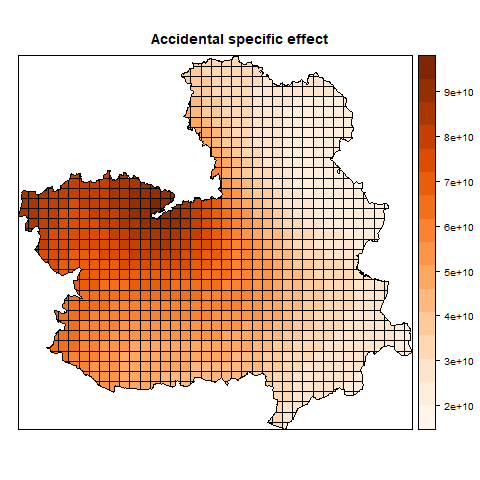}
\includegraphics[scale=0.40]{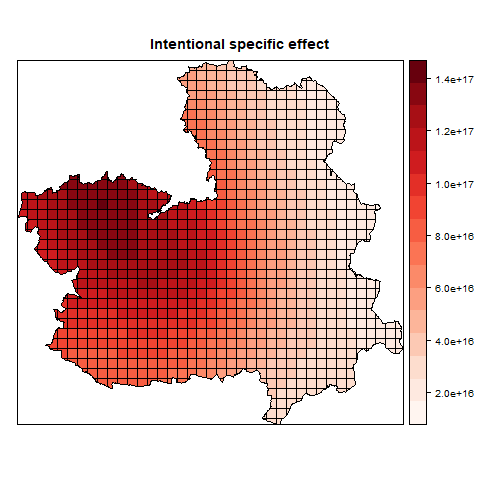}
\includegraphics[scale=0.40]{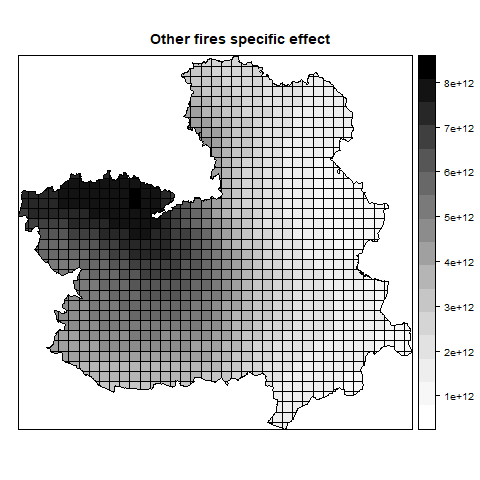}
\end{center}
\caption{Posterior mean of the shared spatial effect (top-left), the accidental-specific (top-right), the intentional-specific (bottom-left) and the other-specific (bottom-right) spatial effects.}
\label{fig:clmfires:est2}
\end{figure}

Figure~\ref{fig:clmfires:est1} shows the posterior mean of the log-intensity of each type of the forest fires where a really different spatial pattern can be seen for each type. Specifically, lightning fires are situated in the east part of region, while the other three types of forest fires are more likely to appear in the west and the center part of the region. Additionally, the posterior means of the shared and specific spatial effects can also be seen in Figure~\ref{fig:clmfires:est2}. Combining the information of both figures, the values of the specific spatial effects point to the fact that most of the forest fires of the central part of the region are not lightning forest fires.

\section{Discussion}
\label{sec:dis}

\medskip
Multivariate spatial models are widely applied in several fields such as epidemiology or ecology. In this article, we have shown how to apply multivariate spatial models using the \pkg{R-INLA} package and details of how to analyse each spatial data type (lattice, geostatistics and point patterns) have been specified. Furthermore, we have illustrated the application of these models using three available datasets: \code{clmfires} dataset in the \pkg{spatstat} package, the \code{meuse} dataset in the \pkg{gstat} package and simulated data available from \url{https://github.com/becarioprecario/INLAMCMC_spatial_examples/tree/master/dismap_example}. The idea of illustrating our objective with straightforward examples was the main cause of the choose of these models. However, more complex spatial and spatio-temporal structures can be proposed and, obviously, \pkg{R-INLA} is able to adjust more complex spatial and spatio-temporal models.

\medskip
The main goal of this work has been to illustrate how to perform multivariate spatial Bayesian inference using \pkg{R-INLA}. Other alternatives based on MCMC algorithms may be highly computational demanding in a multivariate spatial context, therefore, becoming unwise in analysing larger spatial datasets. Hence, it has been shown that \pkg{R-INLA} is a interesting and a worthwhile toolbox to apply multivariate spatial models. Additionally, the necessary R scripts to reproduce the examples are published in \url{https://github.com/FranciscoPalmiPerales/Mult-Sp-INLA}. 

\medskip
In conclusion, the present work explains how to perform Bayesian multivariate spatial inference using \pkg{R-INLA}. We have exemplified how to perform these analysis using public datasets paying special attention to the main steps of the code.

\section{Acknowledgements}

V. G\'omez-Rubio has been supported by grants SBPLY/17/180501/000491 and SBPLY/21/180501/000241, funded by Consejer\'ia de Educaci\'on, Cultura y Deportes (JCCM, Spain) and Fondo Europeo de Desarrollo Regional, grant MTM2016-77501-P and PID2019-106341GB-I00, funded by the Ministerio de Econom\'ia y Competitividad (Spain), and grant PID2019-106341GB-I00, funded by Ministerio de Ciencia e Innovaci\'on (Spain).  F. Palm\'i-Perales was supported by a doctoral scholarship awarded by the University of Castilla–La Mancha (Spain).

\bibliography{RJwrapper}

\address{Francisco Palm\'i-Perales\\
  Department of Applied Economy, Faculty of Economics\\ 
  Universitat de Val\`encia\\
  Av. dels Tarongers, S/N, 46022\\
  Val\`encia, Spain\\
  0000-0002-0751-7315\\
  \email{Francisco.Palmi@uv.es}}

\address{Virgilio G\'omez-Rubio\\
  Department of Mathematics \\
  Universidad de Castilla-La Mancha\\
  Av. de España, s/n, 02001 \\
  Albacete, Spain\\
  0000-0002-4791-3072\\
  \email{Virgilio.Gomez@uclm.es}}

\address{Roger S. Bivand\\
  Department of Economics \\
  Norwegian School of Economics \\
  Helleveien 30, N-5045 \\
  Bergen, Norway \\
  0000-0003-2392-6140\\
  \email{Roger.Bivand@nhh.no}}

\address{Michela Cameletti\\
  Department of Economics.\\ 
  Universit\'a degli studi di Bergamo\\
  Via dei Caniana 2. IT-24127 \\
  Bergamo, Italy\\
  0000-0002-6502-7779\\
  \email{michela.cameletti@unibg.it}}

\address{H\aa{}vard Rue\\
  King Abdullah University of Science and Technology \\
  Thuwal, Saudi Arabia \\
  0000-0002-0222-1881\\
  \email{haavard.rue@kaust.edu.sa}}


\end{article}

\end{document}